\documentclass[doublecol]{epl2} 
\usepackage{graphicx}
\usepackage{float}
\usepackage{dcolumn}
\usepackage{float}
\usepackage{bm}
\usepackage{mathrsfs}
\usepackage{ulem}
\usepackage{hyperref}
\usepackage{xcolor}
\usepackage{soul}
\usepackage{amssymb}
\usepackage{amsthm}
\usepackage{mathtools}

\title{Topological nature of the transition between the gap and the gapless superconducting states}

\author{Yuriy Yerin\inst{1} \and A.A.~Varlamov\inst{2} \and Caterina Petrillo\inst{1}}

\institute{                    
  \inst{1} Dipartimento di Fisica e Geologia, Universitá degli Studi di Perugia - Via Pascoli, 06123 Perugia, Italy\\
  \inst{2} CNR-SPIN, via del Fosso del Cavaliere, 100, 00133 Roma, Italy }

\abstract{Recently it was demonstrated that the long-known transition between the gap and gapless superconducting states in the Abrikosov-Gor'kov theory of superconducting alloy with paramagnetic impurities is of the Lifshitz's type, i.e. at zero temperature this is the $2\frac12$ order phase transition. Since transitions of this kind in a normal metal are always associated to certain topological changes, then below we clarify the topological nature of the transition under consideration. Namely, we demonstrate that the topological invariant which in process of the transition undergoes the change is nothing but the Euler characteristic.  Alternatively, in terms of the theory of catastrophes one can relate this transition to appearance of the cuspidal edge at the corresponding surface of the density of states as the function of energy and superconducting order parameter. The concept of experiments for the confirmation of $2\frac12$ order topological phase transition is proposed. Obtained theoretical results can be applied for the explanation of recent experiments with lightwave-induced gapless superconductivity, for the interpretation of the disorder induced transition $s_{\pm}$-$s_{++}$ states via gapless phase in two-band superconductors, and the emergence of gapless color superconductivity in quantum chromodynamics.}

\begin{document}

\maketitle

\textbf{Introduction}. - The study of topological phase transitions is becoming one of a hot research areas in the condensed matter physics. The term {\it topological phase transition} firstly appeared in the condensed matter being related to the Lifshitz's transitions \cite{Lifshitz1960, Volovik1, Volovik2, Varlamov1}, where the {\it number of the components of topological connectivity} of the Fermi surface (FS) undergoes changes under the effect of different factors: pressure, magnetic field, doping, etc. In the thermodynamic description such a transition is manifested by the specific square root singularity in the third derivative of the free energy over the parameter governing transition. It is why, according to the Ehrenfest terminology \cite{Jaeger1998}, the latter was labeled by I.M. Lifshitz as the phase transition of fractional, $2\frac12$, order. The Lifshitz transition leads to observable anomalies in the electron characteristics of metals such as heat capacity and conductivity, while the Seebeck coefficient experiences the giant singularity \cite{Watlington, Egorov1983, Varlamov1989, Blanter1994}.

Nowadays there are a plenty of exotic materials that exhibit nontrivial topological properties: topological insulators, topological  superconductors, topological superfluids, Dirac and Weyl semimetals, etc. To this end a whole zoo of topological invariants such as winding numbers, Chern numbers, Betti numbers and Euler characteristics is exploited to quantify and classify different phases of matter and topological transitions between them \cite{Simon}. The important hallmark of all mentioned above topological invariants characterizing this or that transition is their insensitivity to smooth deformations of the phase space on which they are defined.
Consequently, a topological phase transition between two distinct phases can be described as a change in the topological invariant at the transition point. Along with that from the physical point of view, topological transitions are remarkable by the fact that they are usually accompanied by the collapse of a gap in the energy spectrum of the system.

It is all the more surprising that, against the background of such complex phenomena, it was recently discovered \cite{Yerin_scipost} that the transition between the well-known gap and gapless states of a  superconductor with magnetic impurities belongs to the Lifshitz type, i.e. has the character of the  $2\frac12$ phase transition, and therefore, one should expect that some topological changes can be also associated with it. 

As a matter of fact the phenomenon of gapless superconductivity was predicted by Abrikosov and Gor'kov (AG) as a result of the extension of the  theory of  superconducting alloys\cite{AG1959} to the case of paramagnetic impurities \cite{AG1960}. One of important consequences of the AG theory was the statement that the initial identification of the supercurrent flow with the presence of a gap in the quasiparticles spectrum by the authors of the BCS theory was too restrictive.

 The transition between gap and gapless regimes is driven by the concentration of paramagnetic impurities in the frameworks of the original AG theory \cite{AG1960, Ambegaokar, Maki1968}. Gapless superconductivity occurs in the very narrow interval of impurity concentrations $0.912\,n_c <n <n_c$, where $n_c$ is the concentration that completely suppresses the supercurrent flow. Later it was recognized that the gapless regime in a superconductor can be induced also by different mechanisms breaking the time-reversal symmetry: magnetic field \cite{Maki1968}, flowing current itself \cite{Maki1968}, proximity effect \cite{Hauser} and the light \cite{Yang2019}.

To clarify the topological nature of the transition between the gap and gapless states, we propose an alternative (with respect to the fundamental article \cite{Lifshitz1960}) view at the problem by studying the topological evolution of the surface of the quasiparticle density of states (DOS) $N(\omega,\Delta_0)$ as a function of the paramagnetic impurities concentration in superconductor. We choose the energy ($\omega$) and the value of the order parameter ($\Delta_0$) of the superconductor in the absence of impurities as the phase space for determining this function. We will demonstrate that the Euler-Poincar{\`e} characteristic  undergoes the jump in the transition point between the gap and gapless states. 

Finally, we propose a strategy for the experimental confirmation and the verification of the topological nature of such a transition. 

\textbf{Free energy}. - The expression for the free energy in the vicinity of the transition between gapless and gap regimes at $T=0$ (see \cite{Maki1968, Skalski1964}) is given by
\begin{equation}
\label{free_energy}
{F_{s - n}}\! = \! - \frac{{N\left( 0 \right){{\Delta }^2}}}{2}\left\{ \begin{gathered}
  1 - \frac{\pi }{2}\zeta  + \frac{2}{3}{\zeta ^2},{\text{ }}\zeta  \leqslant 1, \hfill \\
  1\! - \zeta \arcsin {\zeta ^{ - 1}} \!+\! {\zeta ^2}\left( {1\! -\! \sqrt {1\! -\! {\zeta ^{ - 2}}} } \right) \hfill \\
   - \frac{1}{3}{\zeta ^2}\left( {1 - {{\left( {1 - {\zeta ^{ - 2}}} \right)}^{3/2}}} \right),{\text{ }}\zeta  > 1,{\text{ }} \hfill \\
\end{gathered}  \right.
\end{equation}
\noindent
where $\Delta=\Delta(\tau_s)$ is the order parameter in the presence of impurities ($\Delta  \in \mathbb{R}$) and $N(0)=\frac{mp_F}{\pi^2 \hbar^3}$ is the density of states (DOS) at the Fermi level. The parameter 
\begin{equation}
\label{zeta}
\zeta  = (\tau _{s}\Delta)^{-1}, 
\end{equation}
\noindent
with $\tau_{s}$ as the electron spin-flip scattering time due to the presence of paramagnetic impurities, governs the transition between the gap and gapless states.  For $0 < \zeta < 1$ the energy gap $\Delta_g$ in the  quasiparticle spectrum of superconductor has a nonzero value, while for $\zeta \geq 1$   $\Delta_g \equiv 0$:
the gapless state is realized. At the same time the value of order parameter $\Delta$ in this regime remains different from zero and the phenomenon of supercurrent flow occurs. The critical point $\zeta = 1$  separates the gap and the gapless states. 
\begin{figure}
\includegraphics[width=1\columnwidth]{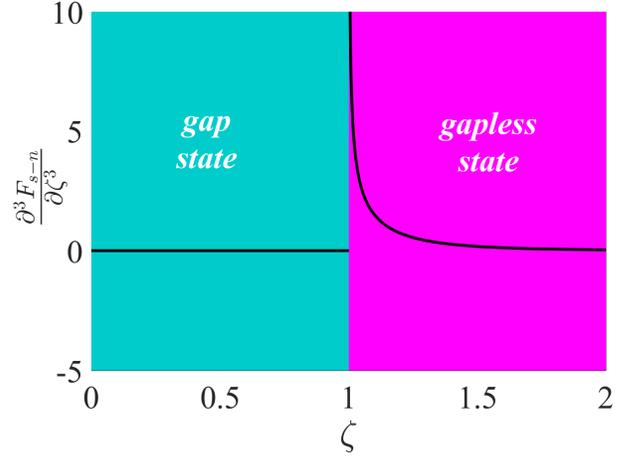}
\caption {The third derivative of the free energy Eq. (\ref{free_energy}) with respect to $\zeta$ given by Eq. (\ref{3d_free_energy}). The essential discontinuity is clearly observed at $\zeta=1$.  Cyan and magenta colors in the background of the plot separate gap and gapless states respectively.}
\label{3d_derivative_energy}
\end{figure}

To make sure that the gap-gapless transition is the Lifshitz phase transition of the $2\frac12$ order it is enough to calculate the third derivative of the free energy (Eq. (\ref{free_energy})) over the parameter $\zeta$ that drives the transition. One can see that the first and the second derivatives remain continuous functions at $\zeta=1$, while the third derivative 
\begin{equation}
\label{3d_free_energy}
\frac{{{\partial ^3}{F_{s - n}}}}{{\partial {\zeta ^3}}} = N\left( 0 \right){\Delta ^2}\left\{ \begin{gathered}
  0,{\text{ }}\zeta  \leqslant 1, \hfill \\
  \frac{1}{{{\zeta ^4}\sqrt {{\zeta ^2} - 1} }},{\text{ }}\zeta  > 1,{\text{ }} \hfill \\ 
\end{gathered}  \right.
\end{equation}
\noindent
clearly shows the essential discontinuity with the square root singularity from the gapless side (see Fig. \ref{3d_derivative_energy}). This behavior is completely analogous to the Lifshitz $2 \frac12$ order phase transitions in metals. The similarity is also confirmed by the quasiparticle DOS dependence on the parameter $\zeta$. 
It was shown \cite{Maki1968, Ambegaokar, Skalski1964} that the quasiparticle DOS of a superconductor $N_s\left( \omega  \right)$ is finite at $\omega=0$ and has a typical cusp for $2\frac12$ order phase transition at $\zeta=1$
\begin{equation}
N_s\left( 0  \right) = N\left( 0 \right)\frac{\sqrt{\zeta+1}}{\zeta}\sqrt{\zeta-1}.
\end{equation}

\textbf{Topological interpretation}. 
- So we made sure that the transition between the gap and gapless states in superconductor has the nature  of the Lifshitz transition, however its topological interpretation is missing.  In the case of a normal metal the latter is trivial: the topological modification of the FS occurs when the chemical potential $\mu$ reaches a certain critical value $\mu_c$. In result the number of components of  topological connectivity of the FS changes. At this point the  parameter $z=\mu-\mu_c$ governing the transition passes through its zero value \cite{Blanter1994, Volovik1, Volovik2, Varlamov1}. 
\begin{figure*}
\includegraphics[width=0.67\columnwidth]{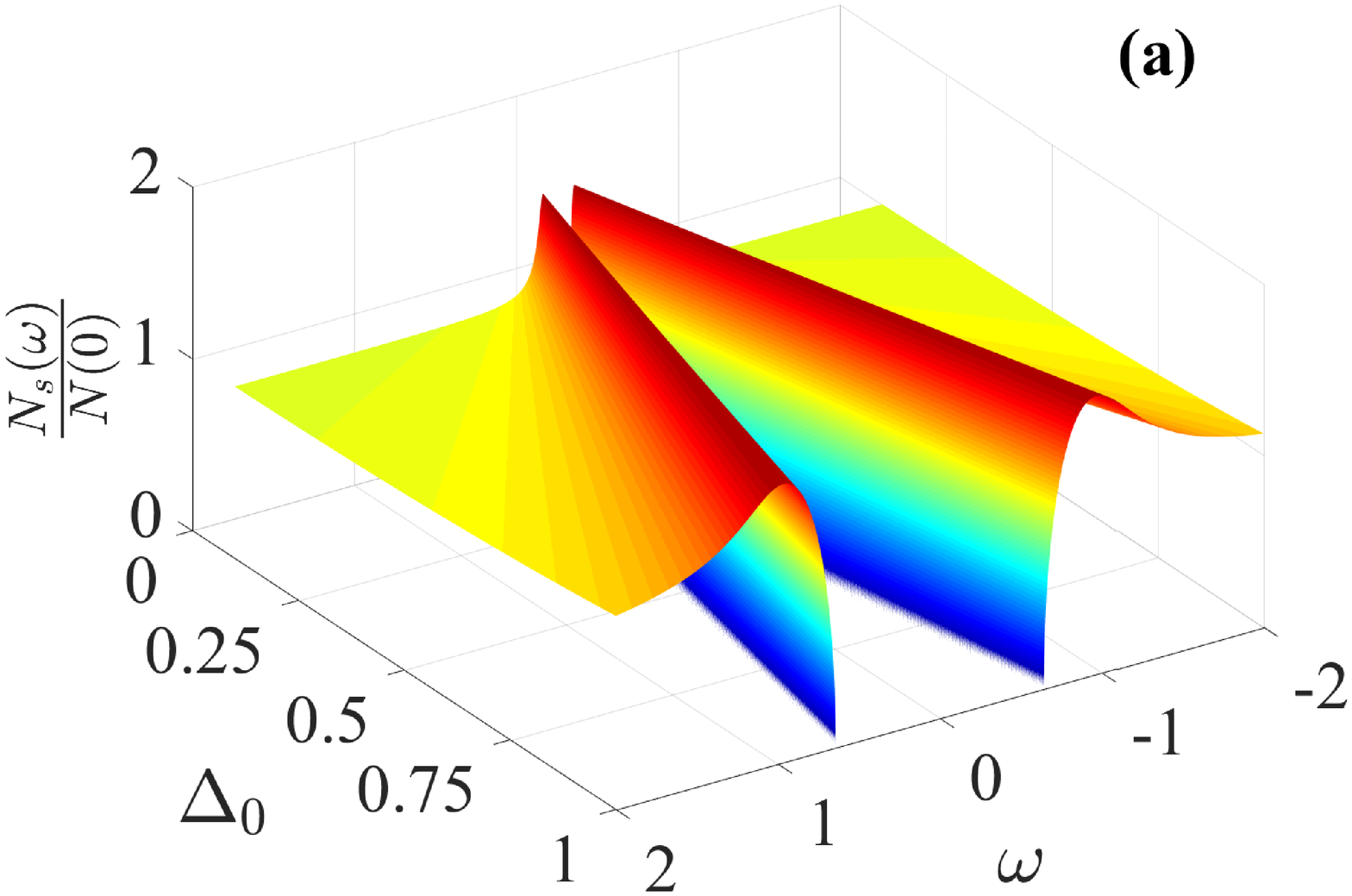}
\includegraphics[width=0.67\columnwidth]{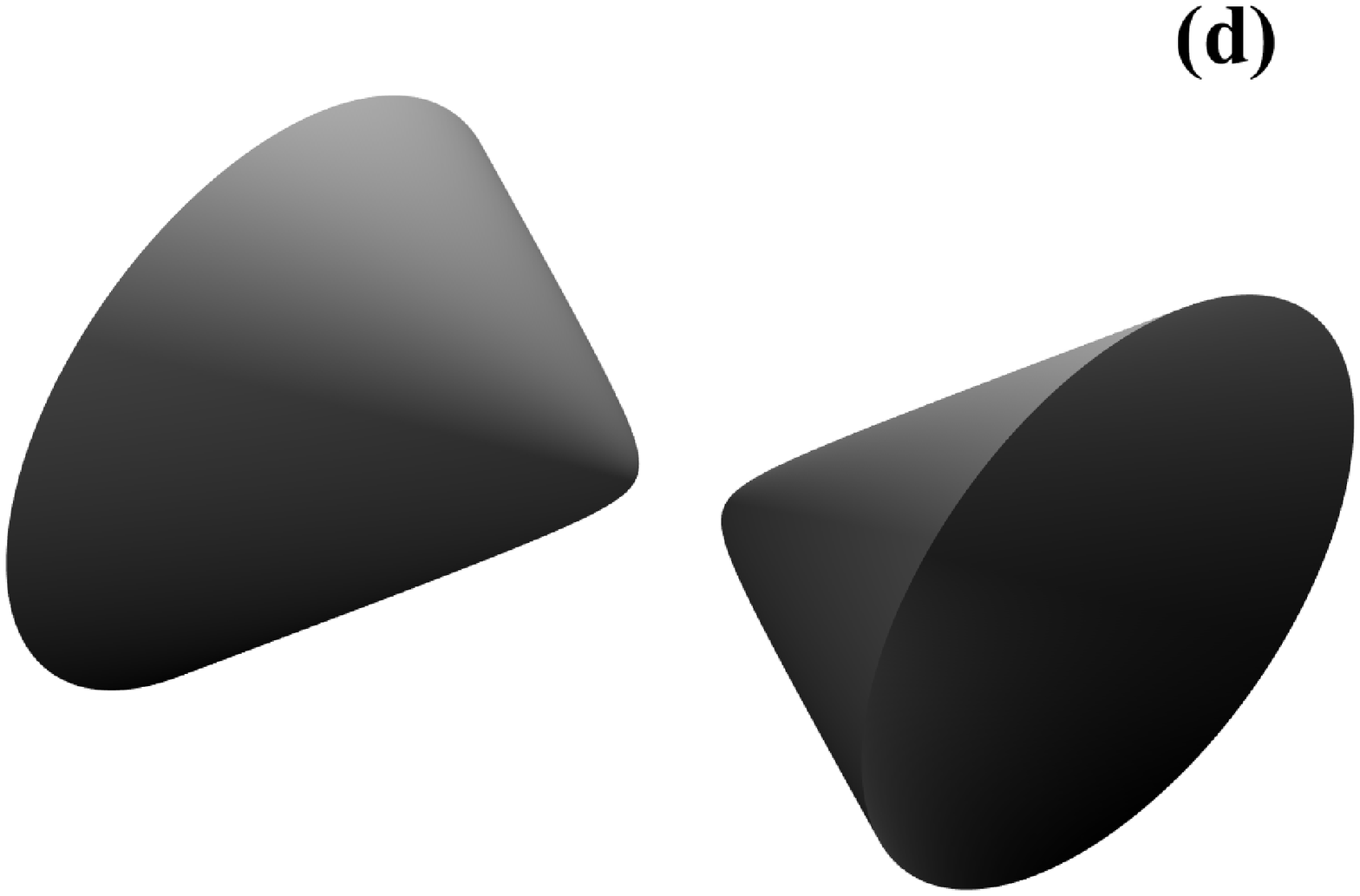}
\includegraphics[width=0.67\columnwidth]{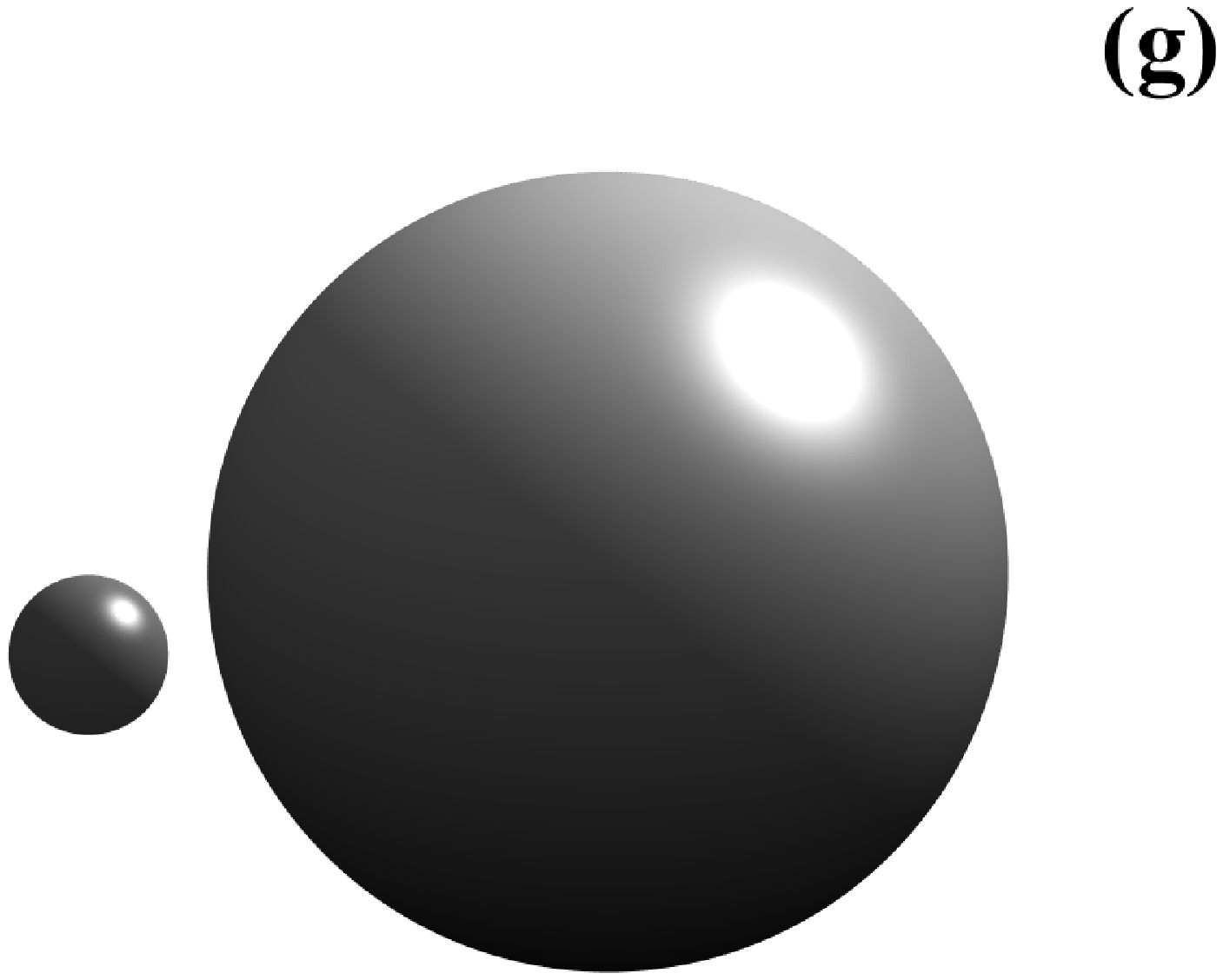}
\includegraphics[width=0.67\columnwidth]{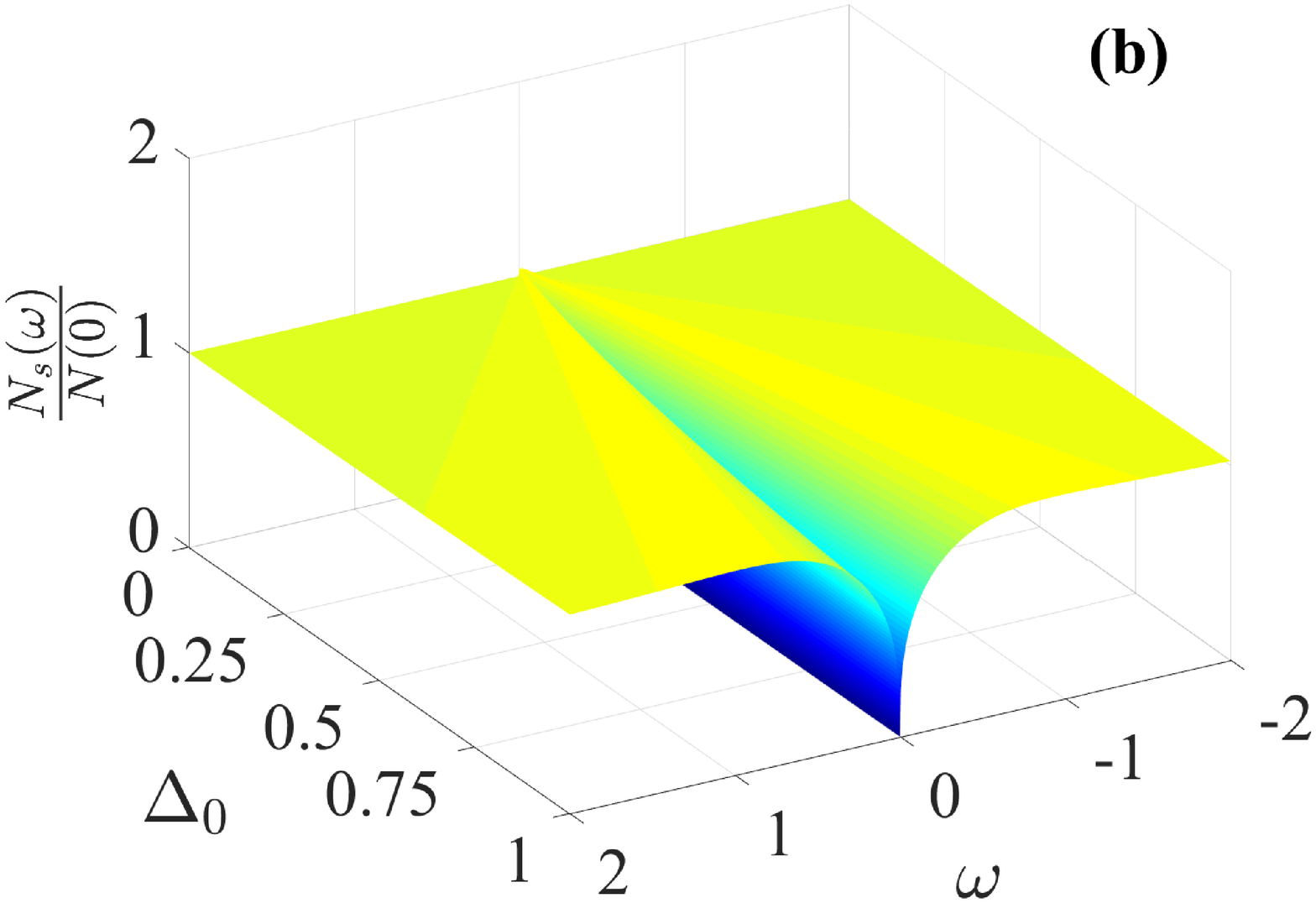}
\includegraphics[width=0.67\columnwidth]{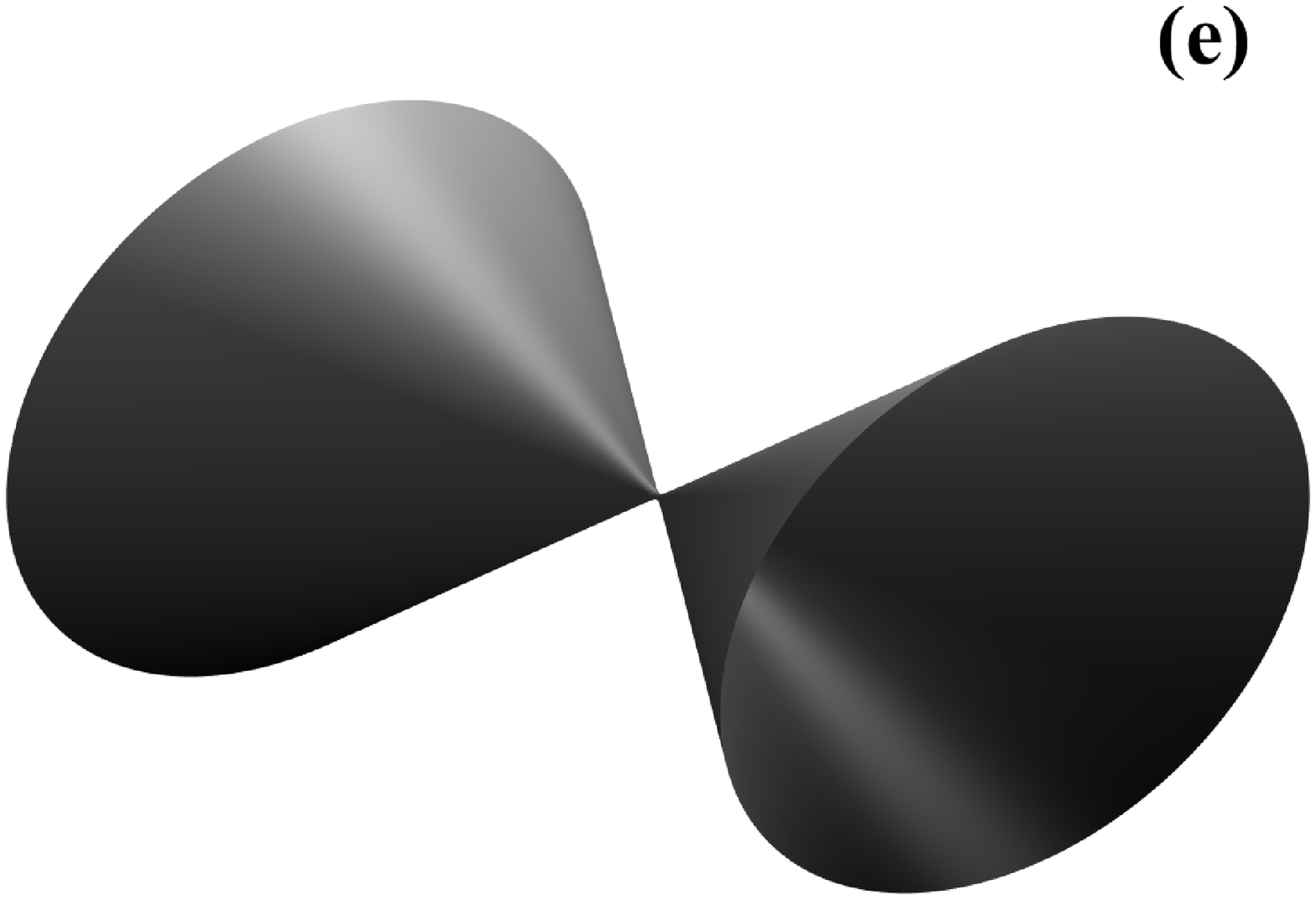}
\includegraphics[width=0.67\columnwidth]{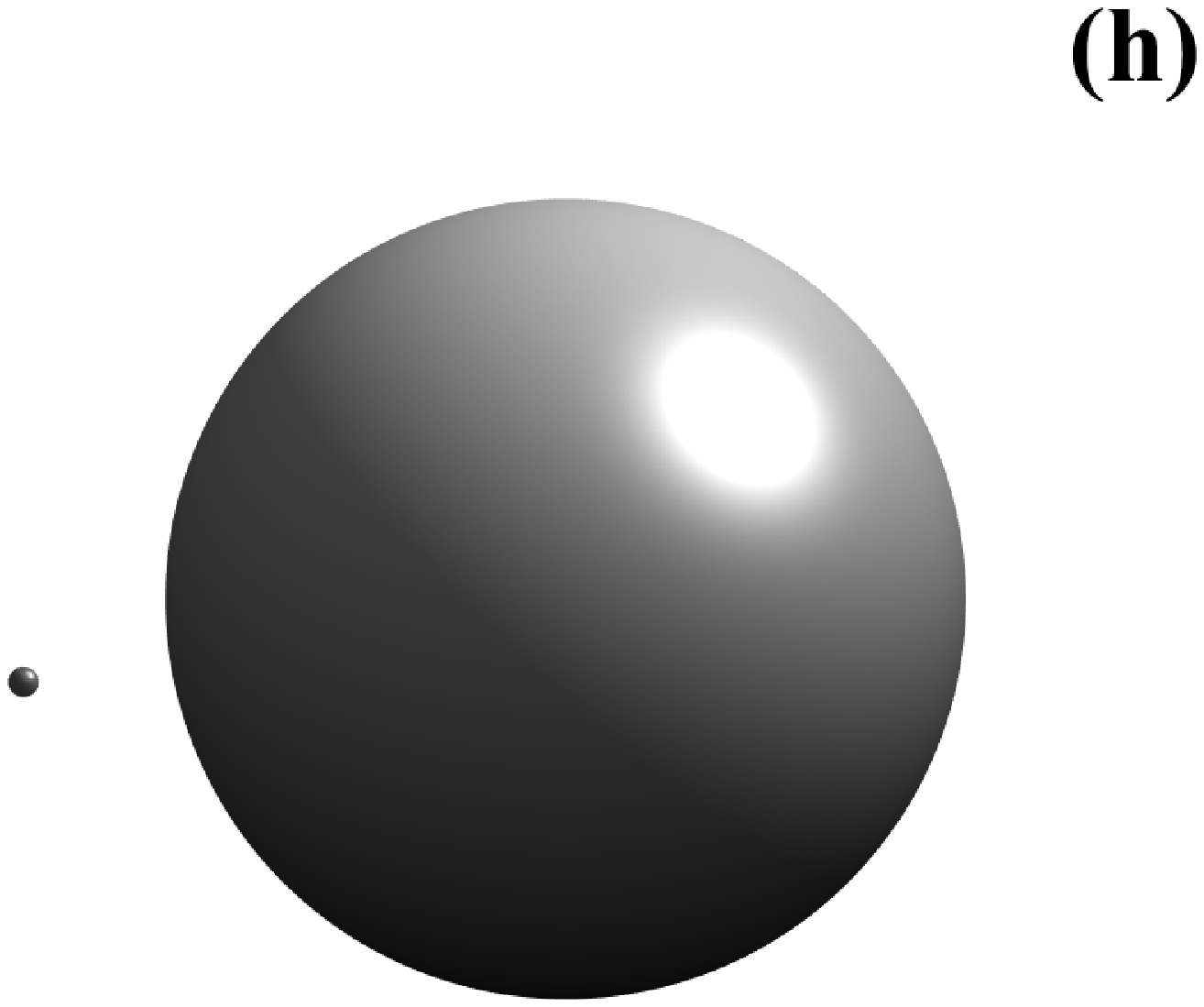}
\includegraphics[width=0.67\columnwidth]{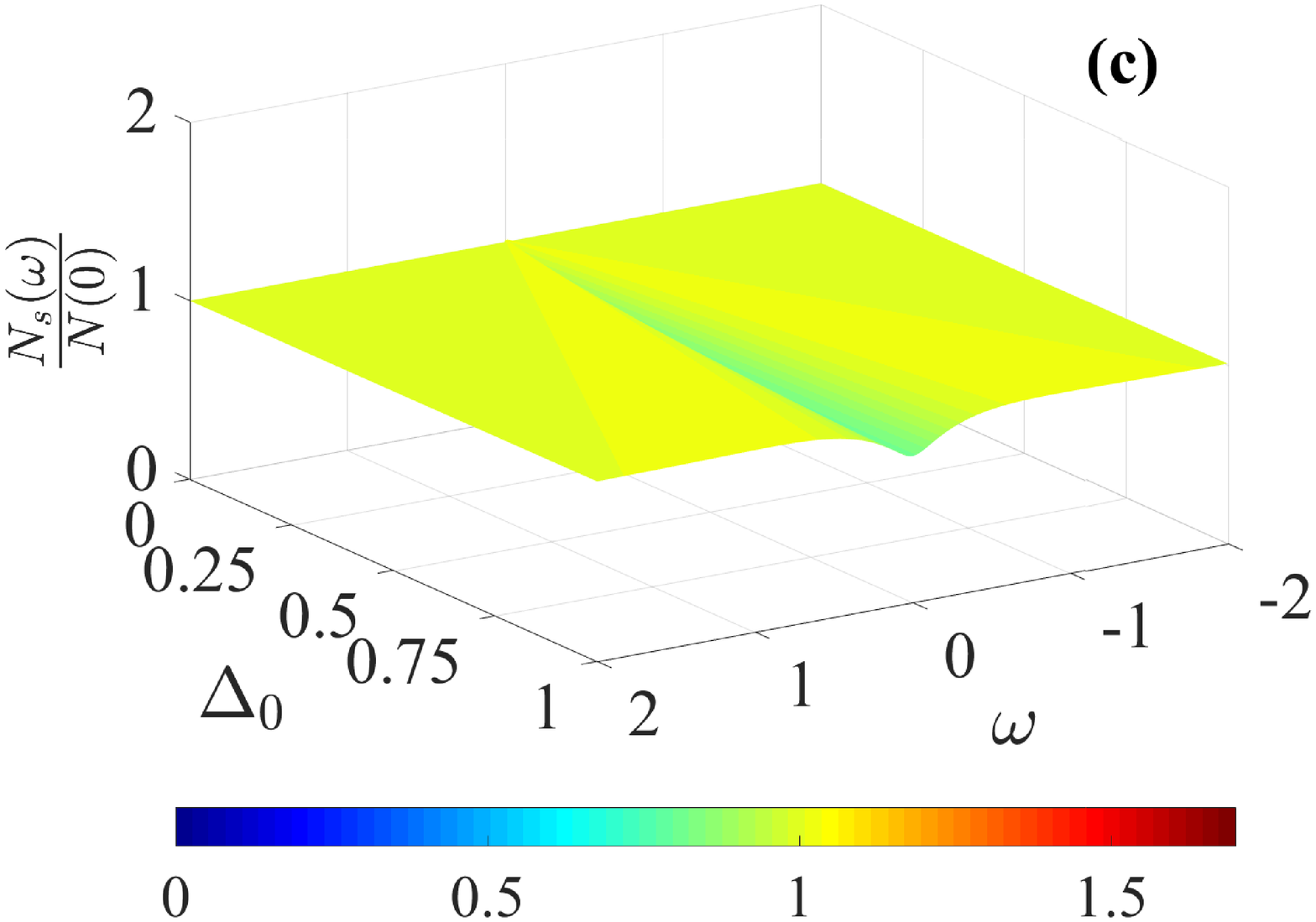}
\includegraphics[width=0.67\columnwidth]{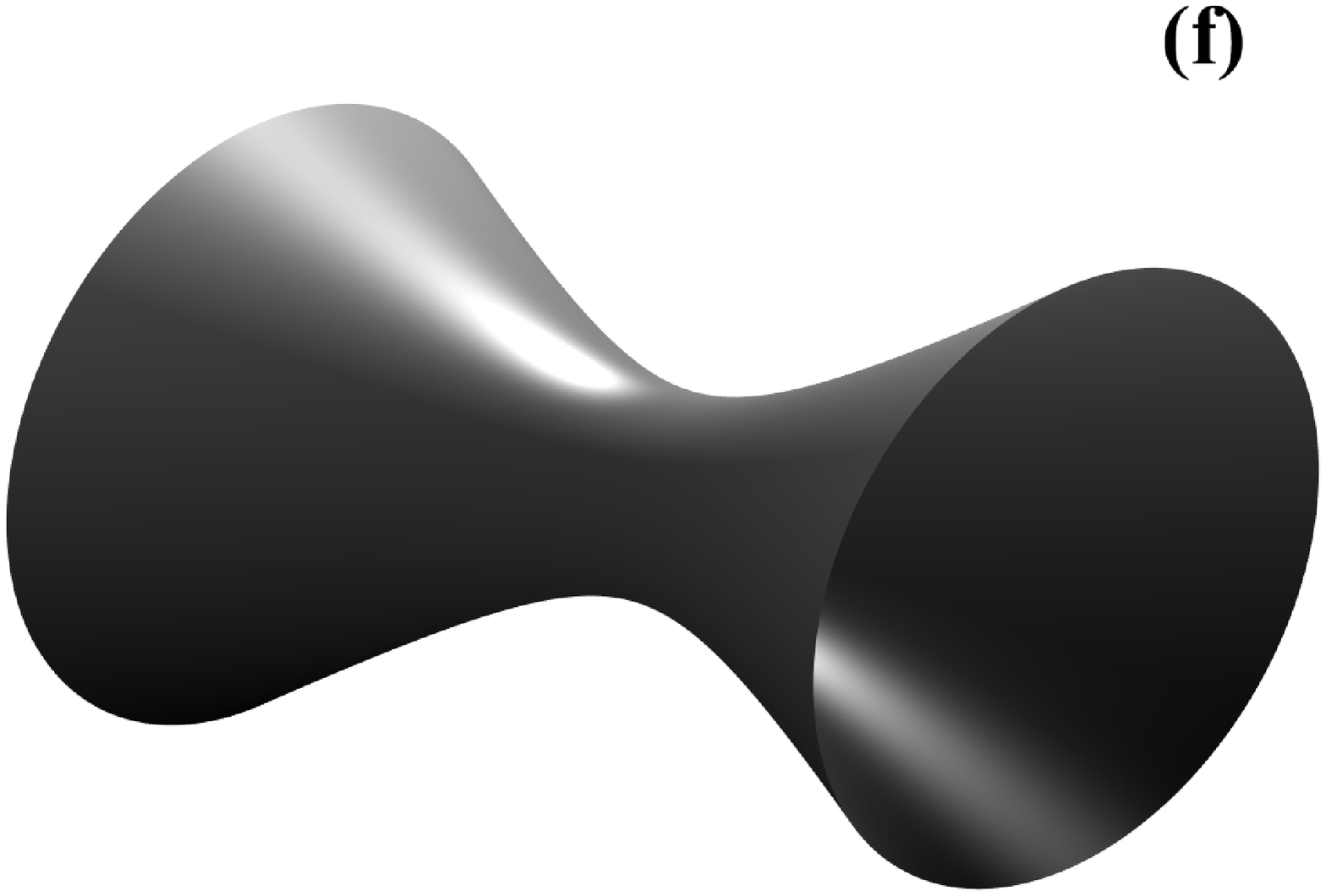}
\includegraphics[width=0.67\columnwidth]{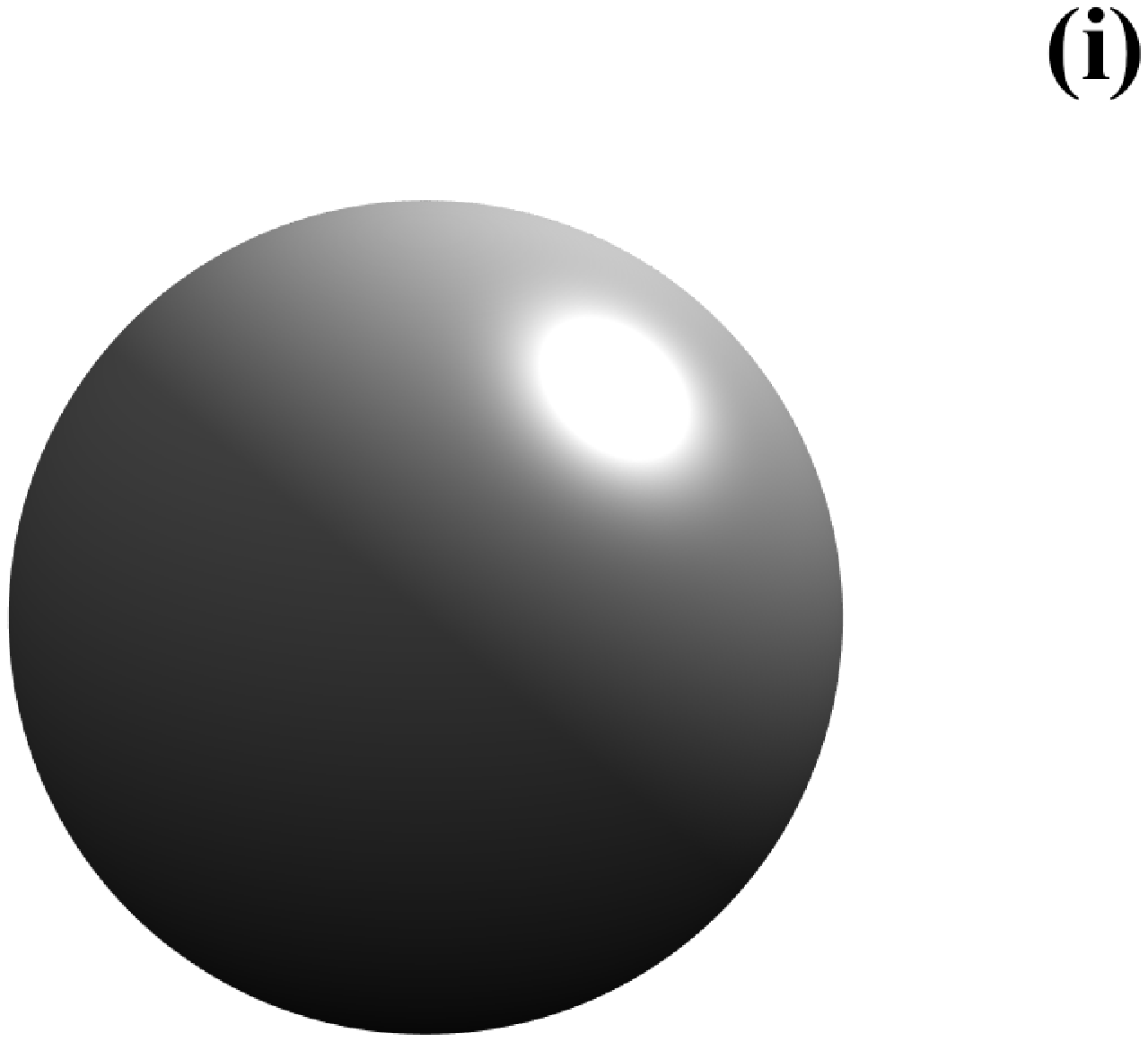}
\caption{Topological evolution of the quasiparticle DOS in the $\omega$-$\Delta_0$ phase space (a-c) combined with the similar evolution of the FS in the momentum space for the Lifshitz transition (d-i). Figure (a) with $\zeta=0.1$ is topologically equivalent to figure (d) and (g) with $z<0$; figure (b) with $\zeta=1$ is topologically equivalent to figure (e) and (h) with $z=0$ and figure (c) with $\zeta=2$  is topologically equivalent to figure (f) and (i) with $z>0$. For a better understanding of the DOS topology, a video is also available at \cite{video}.}\label{DOS_topology}
\end{figure*}

At this point we should note that the  FS does not manifest itself directly in the phenomenon of superconductivity. At the same time in a metal undergoing the Lifshitz transition side by side with the topological change of the FS a gap appears in the quasiparticle DOS energy dependence. This fact should be manifested in the topological properties of the corresponding surfaces (yet, to the best of our knowledge these changes were not classified). It is why below we turn to study of the topological properties of the quasiparticle DOS surface in the phase space of $\omega$-$\Delta_0$ based on the general expression for $N\left( \omega, \Delta_0 \right)$ \cite{AG1960, Ambegaokar}
\begin{equation}
\label{DOS_general}
N\left( \omega, \Delta_0 \right) = N\left(0, \Delta_0 \right){\zeta ^{ - 1}}\operatorname{Im} u,
\end{equation}
where $u$ is given by the implicit function
\begin{equation}
\label{u}
\frac{\omega }{{\Delta}} = u\left( {1 - \frac{\zeta }{{\sqrt {1 - {u^2}} }}} \right),
\end{equation}
 and the expression for the order parameter $\Delta$ at $T=0$ \cite{AG1960, Maki1968}
\begin{equation}
\label{OP_T=0}
\ln \left( {\frac{\Delta }{{{\Delta _0}}}} \right) \!= \!\left\{ \begin{gathered}
  - \frac{\pi }{4}\zeta ,{\text{ }}\zeta  \leqslant 1, \hfill \\
   - \operatorname{arcosh} \zeta \!  - \! \frac{1}{2}\left( {\zeta \arcsin {\zeta ^{ - 1}} - \sqrt {1 - {\zeta ^{ - 2}}} } \right),{\text{ }} \hfill \\
  \zeta  > 1, \hfill \\ 
\end{gathered}  \right.
\end{equation}
where we recall that $\Delta_0 \neq 0$ is the value of the superconducting order parameter in the absence of magnetic impurities.

Three topologically dissimilar surfaces for $\zeta <1$, $\zeta=1$ and $\zeta >1$ are shown in Figure  \ref{DOS_topology}a-c respectively. Figure \ref{DOS_topology}a corresponds to the gap state with $0 < \zeta < 1$ and with the characteristic narrowing ``valley'' between two sheets of the DOS surfaces. When $\zeta=1$ the collapse of the energy gap occurs with the formation a topological feature known as the cuspidal edge \cite{Arnold1} at $\omega=0$ (Fig. \ref{DOS_topology}b), indicating  the emergence of the catastrophe phenomenon in the $\omega$-$\Delta_0$ space over the gap-gapless phase transition \cite{Arnold2,  Saji}. Figure \ref{DOS_topology}c corresponds to the gapless state ( $\zeta>1$) and exhibits the gradual degradation of the DOS curved surface to a plane as $\zeta  \to \infty$. In this representation one can freely ``travel'' over the each surface $N(\omega, \Delta_0)$ by changing the variables $\omega$ and $\Delta_0$ while keeping $\zeta=\zeta(\Delta_0, \tau_s)=$const and adjusting the value of $ \tau_s$ for each $\Delta_0$ to satisfy the constancy of the given value of $\zeta$. 
The manipulation of $\tau_s$ for each $\Delta_0$ in order to carry $\zeta = $ const  for the given DOS surface is not artificial procedure. In some sense during the study of the Lifshitz transition the experimentalists do the same, for example, during the investigation of the anomalous behavior of thermopower in ${\text{L}}{{\text{i}}_{{\text{1 - x}}}}{\text{M}}{{\text{g}}_{\text{x}}}$ alloy in a dependence of $x$ close to transition at $x =$ 0.19  \cite{Egorov1983}. 
Corresponding movies with the rotated DOS surfaces were created for visualization of their topology \cite{video}.

The subsequent topological interpretation of the gap-gapless transition can be performed by introducing the topological invariant called the Euler (or Euler-Poincar\'e) characteristic $\chi$ to describe the evolution of the function $N(\omega, \Delta_0)$ (see Figure  \ref{DOS_topology}a-c).  Generally speaking, the Euler characteristic is determined by the integral of the Gaussian curvature over the whole surface via the Gauss-Bonnet theorem \cite{Spanier1966}. The Gaussian curvature can be calculated with use of asymptotic expressions for $N_s\left( \omega, \Delta_0 \right)$ for the gap state ($0<\zeta < 1, \Delta_g \ne 0$) \cite{Maki1968} 
\begin{equation}
\label{DOS_gap}
N_s\left( \omega, \Delta_0  \right) \!= \! N\left( 0 \right)\!\left\{ \begin{gathered}
  0,{\text{ }}\omega  < \Delta_g,  \hfill \\
  {\zeta ^{ -\frac23}}{\left( {1\!-\!{\zeta ^{\frac23}}} \right)^{ - \frac14}}\!\sqrt {\frac{2}{3}\frac{{\omega \!-\!\Delta_g }}{\Delta }} ,{\text{ }}\omega  \geqslant \Delta_g, \hfill \\ 
\end{gathered}  \right.
\end{equation}
for the gapless state ($\zeta > 1, \Delta_g = 0$)
\begin{equation}
\label{DOS_gapless}
\begin{gathered}
  {N_s}\left( {\omega , \Delta_0} \right) = N\left( 0 \right)\left[ {{{\left( {1 - {\zeta ^{ - 2}}} \right)}^{1/2}}} \right. \hfill \\
  \left. { + \frac{3}{2}{\zeta ^{ - 4}}{{\left( {1 - {\zeta ^{ - 2}}} \right)}^{ - 5/2}}{{\left( {\frac{\omega }{{\Delta}}} \right)}^2}} \right], \hfill \\ 
\end{gathered}
\end{equation}
and for the special case of $\zeta=1$
\begin{equation}
\label{DOS_gapless_zeta=0}
{N_s}\left( {\omega ,\Delta_0} \right) = N\left( 0 \right)\frac{{\sqrt 3 }}{2}\left[ {{{\left( {\frac{{2\omega }}{\Delta }} \right)}^{1/3}} - \frac{1}{{24}}{{\left( {\frac{{2\omega }}{\Delta }} \right)}^{5/3}}} \right],
\end{equation}
to avoid a singularity problem with the divergence at ``suspicious'' line $\zeta=1$ in Eq. (\ref{DOS_gapless}).

Being a function of two variables Eqs. (\ref{DOS_gap})-(\ref{DOS_gapless_zeta=0}) allow to evaluate the Gaussian curvature at the given point of the surfaces by means of the formula
\begin{equation}
\label{curvature}
K = \frac{{\frac{{{\partial ^2}N}}{{\partial {\omega ^2}}}\frac{{{\partial ^2}N}}{{\partial {\Delta_0 ^2}}} - {{\left( {\frac{{{\partial ^2}N}}{{\partial \omega \partial \Delta_0 }}} \right)}^2}}}{{{{\left( {1 + {{\left( {\frac{{\partial N}}{{\partial \omega }}} \right)}^2} + {{\left( {\frac{{\partial N}}{{\partial \Delta_0 }}} \right)}^2}} \right)}^2}}},
\end{equation}
and after that to take a surface integral according to the Gauss-Bonnet theorem
\begin{equation}
\label{GB_theorem}
\int\limits_\Omega  {Kd\sigma  = 2\pi \chi },
\end{equation}
where the integration are carried out over the DOS surface ${\Omega}$ and where $d\sigma$ is the element of area of ${\Omega}$.

Based on Eqs. (\ref{curvature}) and (\ref{GB_theorem}) one can find that during the topological transformation throughout the gap-gapless phase transition, the Euler characteristic changes from $\chi=2$ (gap state) to $\chi=1$ (gapless state). We follow the numerical procedure for the calculation of the surface integral Eq. (\ref{GB_theorem}) described in detail in \cite{BoLiu} (see chapter 5.3.1 therein adopted for 2D case) with the implementation by means of Matlab.
\footnote{It may seem that the Euler characteristic for the DOS surface corresponding of the gap collapse  ($\zeta=1$, see Figure \ref{DOS_topology}b) should intuitively take a fractional value between 1 and 2. Generally speaking, topology admits an existence of geometrical objects, known as orbifolds, with fractional values of the Euler characteristic. This is especially relevant for two-dimensional orbifolds \cite{Thurston, Atiyah, Hirzebruch}. In this case we need to apply a generalized formula for the calculation of the Euler characteristic, which takes into account the presence of the nontrivial singular points for the given orbifold such as the corner reflectors, the elliptic points, etc. 
However, despite of the singularity of cuspidal edge and the intuitive expectation of the fractional value of $\chi$, both our numerical calculations based on the Gauss-Bonnet theorem and the naive approach based on the Euler's formula indicate the value of the Euler characteristic  $\chi=1$ for the DOS surface with $\zeta=1$.}

It worth note that one can evaluate the value of  $\chi$ in much more simple way, just performing the polygonization of DOS surfaces and calculating $\chi=V-E+F$ by means of the Euler's formula. Here $V$, $E$, and $F$ are respectively the numbers of vertices (corners), edges and faces of the circumscribed polyhedron \cite{Spanier1966}.

We would like to underline that the numerical evaluation of the Euler characteristic by means of the Gauss-Bonnet theorem was performed to ensure the validity of the obtained result and to compare it with the much easier approach based on Euler's formula.

To illustrate how the concept of the Euler characteristic can be applied to other topological transitions we consider case of the Lifshitz transition and the topological evolution of the FS in the momentum space (Figure \ref{DOS_topology}d-i) . In Figure \ref{DOS_topology}f a one-sheet hyperboloid is not compact. Its deformation retracts onto a circle, and the Euler characteristic is a homotopy invariant, so $\chi = 0$. By tightening the neck of the hyperboloid, it is deformed into a cone as shown in Figure \ref{DOS_topology}e. The cone could be simplexized into 6 singular 2-simplexes giving $\chi = 1$. By detaching the pieces of the cone and smoothing it (Figure \ref{DOS_topology}d), one finds a two-sheet hyperboloid, where each sheet is topologically equivalent to the disc with $\chi = 1$. The Euler characteristic of the disjoint union of two discs is the sum of their Euler characteristics, so $\chi=1+1=2$. Therefore, throughout the Lifshitz transition in Figure  \ref{DOS_topology}d-f the Euler characteristic changes from 2 to 1 and then to 0.

The same interpretation of the Lifshits transition can be done for Figure \ref{DOS_topology}g-i. Here the Euler characteristic changes from $\chi=2$ because of the sphere (Fig. \ref{DOS_topology}i) to $\chi= 2+1=3$, where we consider the additive contributions from the sphere and the point (Fig. \ref{DOS_topology}h). Finally, for Fig. \ref{DOS_topology}g there are two spheres and correspondingly $\chi = 2+2=4$. As a result, we observe the alteration of the Euler characteristic from 4 to 3 and then to 2. 

Therefore, one can conclude that instead of parameter $z$ that governs the Lifshitz transition and controls the corresponding transformation of an open FS into a closed one (Fig. \ref{DOS_topology}d-i) with the emergence of the corresponding gap, the driving parameter for the topological modification under consideration is the value of $(\zeta-1)$. 

We emphasize that the emergence of the nontrivial topology in the form of the cuspidal edge in the gapless state of a superconductor occurs in the $\omega$-$\Delta_0$ space. This phase space differs from the ``traditional'' momentum space applied for the description of gapless states of topological superconductors \cite{Bernevig}. 

Our conclusion remains valid also for finite temperatures $T$ less than the critical temperature $T < T_c$. For $T \ne 0$ the generalized  expression for the free energy $\mathbb{F}_{s - n}$ accounts for the temperature dependence of the order parameter and the Fermi distribution
\begin{equation}
\label{finiteT}
\begin{gathered}
  {\mathbb{F}_{s - n}}\left( {\zeta ,T} \right) = {F_{s - n}}\left( {\zeta ,T} \right) \hfill \\
   - 4T\int\limits_0^\infty  {\ln \left( {1 + \exp \left( { - \frac{\omega }{T}} \right)} \right)} \left( {N\left( {\omega ,{\Delta _0}} \right) - N\left( {0,{\Delta _0}} \right)} \right)d\omega  \hfill \\
   - 2\Delta \left( T \right)N\left( 0 \right)\int\limits_0^\infty  {\frac{{\operatorname{Im} \left( {\frac{1}{{\sqrt {1 - {u^2}} }}} \right)}}{{1 + \exp \left( { - \frac{\omega }{T}} \right)}}} d\omega,  \hfill \\ 
\end{gathered}
\end{equation}
where the first term  functionally remains the same as in Eq. (\ref{free_energy}). We recall that the parameter $u$ is determined by Eq. (\ref{u}). The additive structure of Eq. (\ref{finiteT}) points out that even for finite temperatures the discontinuity of the third derivative is preserved within the superconducting state.

\textbf{Experimental proposals and extensions}. - The experimental verification of the topological phase transition can be performed with a ring, one half of which is a gap superconductor with the concentration of magnetic impurities close to the transition value $\zeta=1$ and the another half is an arbitrary superconductor. In this case, when superconducting contacts have different but close to zero temperature, strong thermoelectric current is induced in the ring and the detected magnetic flux should deviate from the integer values of the magnetic flux quantum $\Phi_0$ \cite{Galperin1973,Galperin1974,Zavaritskii1974,Ginzburg1978,Yerin_scipost}. 
Alternative direction for the study of the gap-gapless phase transition is the measurement of the derivative of the specific heat capacity with respect to the impurity concentration and the observation of the appropriate kink in the dependence.

Our findings can be extended on other physical systems where a gap-closing phenomenon is took place. First, these results may pave a novel way toward to the interpretation of recent experiments with light-wave-driven gapless superconductivity \cite{Yang2019}. It was revealed that  under a lightwave THz radiation supercurrent-carrying states in ${\text{N}}{{\text{b}}_{\text{3}}}{\text{Sn}}$ evolve to long-lived gapless superconductivity with minimal condensate quench.

The second important extension is related to the disorder-induced transition $s_{\pm}$-$s_{++}$ states in two-band superconductors \cite{Barzykin, Efremov, Korshunov, Koshelev}, which are relevant to experimental studies in iron pnictides. According to the theoretical predictions with increasing of the nonmagnetic impurities concentration, one of the gaps is seen to close, leading to a finite residual DOS, followed by a reopening of the gap. The formation of gapless superconductivity as one of the gaps vanishes allows to speculate about the topological nature of $s_{\pm}$-$s_{++}$ transition.

Finally, the topological nature of the gap-gapless transition can be relevant to the phenomenon of gapless color superconductivity in quantum chromodynamics and the string theory \cite{Alford}.  For a color superconductor at zero temperature and at some critical value of the strange quark mass a transition to the gapless color-flavor-locked phase occurs, where the energy gap in the quasiparticle spectrum is not mandatory \cite{Alford, Ruester}.

\textbf{Conclusions}. - We have revealed the topological nature of the transition between the gap and the gapless states of a superconductor. The corresponding topological invariant, namely the Euler characteristic has been applied for the description of the transition.
Moreover, it has been found that this phase transition possesses another interesting topological feature known as the cuspidal edge and related to the catastrophe theory. This allows us to classify this topological phase transition and the collapse of the gap as the catastrophe phenomenon.  
Also, we have proposed several concepts for the experimental confirmation of the $2\frac12$ order topological phase transition. Finally, These results may help to provide new insights into the experiments with lightwave-induced gapless superconductivity and with the disorder induced transition $s_{\pm}$-$s_{++}$ pairing symmetries in two-band superconductors as well as into the theory of gapless color superconductivity in quantum chromodynamics.

\acknowledgments
A.V. is grateful to  Yu.~Galperin, A. I. Buzdin, Aviad Frydman, S. Frolov, S. Bergeret, and O.  Dobrovolskiy for valuable comments. C.P. and Y.Y. acknowledge support by the CarESS project. A.V. acknowledges the financial support under the STSM COST Action CA16218 Nanoscale Coherent Hybrid Devices For Superconducting Quantum Technologies.

\end{document}